\documentclass{article}

\usepackage[accepted]{icml2022}

\usepackage{microtype}
\usepackage{graphicx}
\usepackage{xurl,hyperref}
\usepackage[nolist,nohyperlinks]{acronym}
\usepackage{subcaption}

\begin{acronym}
  \acro{AI}{Artificial Intelligence}
  \acro{ML}{Machine Learning}
  \acro{MAG}{Microsoft Academic Graph}
  \acro{FLOP}{floating point operations}
  \acro{LMs}{language models}
  \acro{PCD}{``Parameter, Compute and Data Trends in Machine Learning''}
\end{acronym}

\icmltitlerunning{Who is leading in AI? An analysis of industry AI research}

\begin{document}

\twocolumn[
\icmltitle{Who is leading in AI? An analysis of industry AI research}

\icmlsetsymbol{equal}{*}

\begin{icmlauthorlist}
    \icmlauthor{Ben Cottier}{epoch}
    \icmlauthor{Tamay Besiroglu}{epoch,futuretech}
    \icmlauthor{David Owen}{epoch}
\end{icmlauthorlist}

\icmlaffiliation{epoch}{Epoch}
\icmlaffiliation{futuretech}{MIT FutureTech}

\icmlcorrespondingauthor{Ben Cottier}{ben@epochai.org}

\vskip 0.3in
]

\printAffiliationsAndNotice{}

\begin{abstract}
\ac{AI} research is increasingly industry-driven, making it crucial to understand company contributions to this field. We compare leading AI companies by research publications, citations, size of training runs, and contributions to algorithmic innovations. Our analysis reveals the substantial role played by Google, OpenAI and Meta. We find that these three companies have been responsible for some of the largest training runs, developed a large fraction of the algorithmic innovations that underpin large language models, and led in various metrics of citation impact. In contrast, leading Chinese companies such as Tencent and Baidu had a lower impact on many of these metrics compared to US counterparts. We observe many industry labs are pursuing large training runs, and that training runs from relative newcomers---such as OpenAI and Anthropic---have matched or surpassed those of long-standing incumbents such as Google. The data reveals a diverse ecosystem of companies steering AI progress, though US labs such as Google, OpenAI and Meta lead across critical metrics.
\end{abstract}

The private sector has come to play a major role in \ac{AI} research and development \cite{ahmed2020dedemocratization,ahmed2023growing}. This rise has been fueled by an explosion in computational resources for AI \cite{sevilla2022compute}, with costs becoming prohibitive for academic labs alone to sustain. The private sector also has an outsized influence: \ac{AI} research articles involving private companies receive about twice as many citations as those without \cite{klinger2022narrowing}. Industry is responsible for almost all of the most compute-intensive training runs \cite{sevilla2022compute}, has a disproportionate influence on systems that are widely deployed, and, as we shall see, develops a large number of the algorithms and architectures that underpin current frontier models.

Recently, the 2023 Stanford \ac{AI} Index \cite{maslej2023aiindex} ranked companies and academic institutions together by their number of publications, and analysed \ac{AI} research at the level of \ac{AI} systems and countries. Among their findings was that Chinese academic institutions have produced the most AI papers in the past decade. In contrast, this work focuses on companies and finds that US tech giants have had a larger research impact than their Chinese counterparts. 

Three datasets inform our findings: \ac{AI} publications from 2010--2023 via OpenAlex, a free and open catalog of scholarly papers \cite{priem2022openalex}, \ac{AI} training compute data from the \ac{PCD} database \cite{epochMachineLearningData2022}, and a novel publicly available dataset of key algorithmic innovations underpinning large language models. Using these datasets we compare leading companies by publications, citations, number of unique authors, largest training runs, and the adoption of their algorithmic innovations. Together, these metrics provide a more holistic view of industry \ac{AI} research compared to previous work. Our results characterize how companies have influenced the field in the past, and suggest how this may continue into the future. This is valuable for AI policy discussions concerning the most important players in the AI industry.

\section*{Results}

\begin{figure*}[t!]
    \centering
    
    \begin{subfigure}{0.49\linewidth}
        \includegraphics[width=\textwidth]{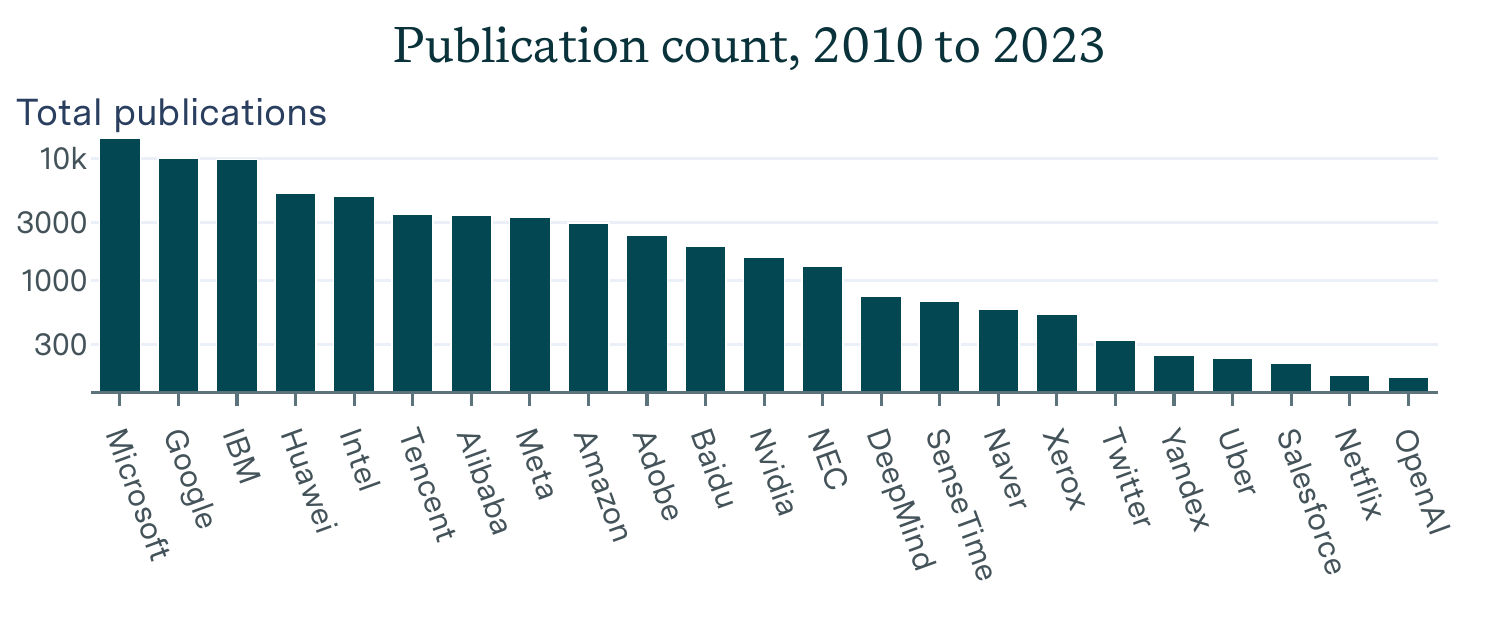}
        \includegraphics[width=\textwidth]{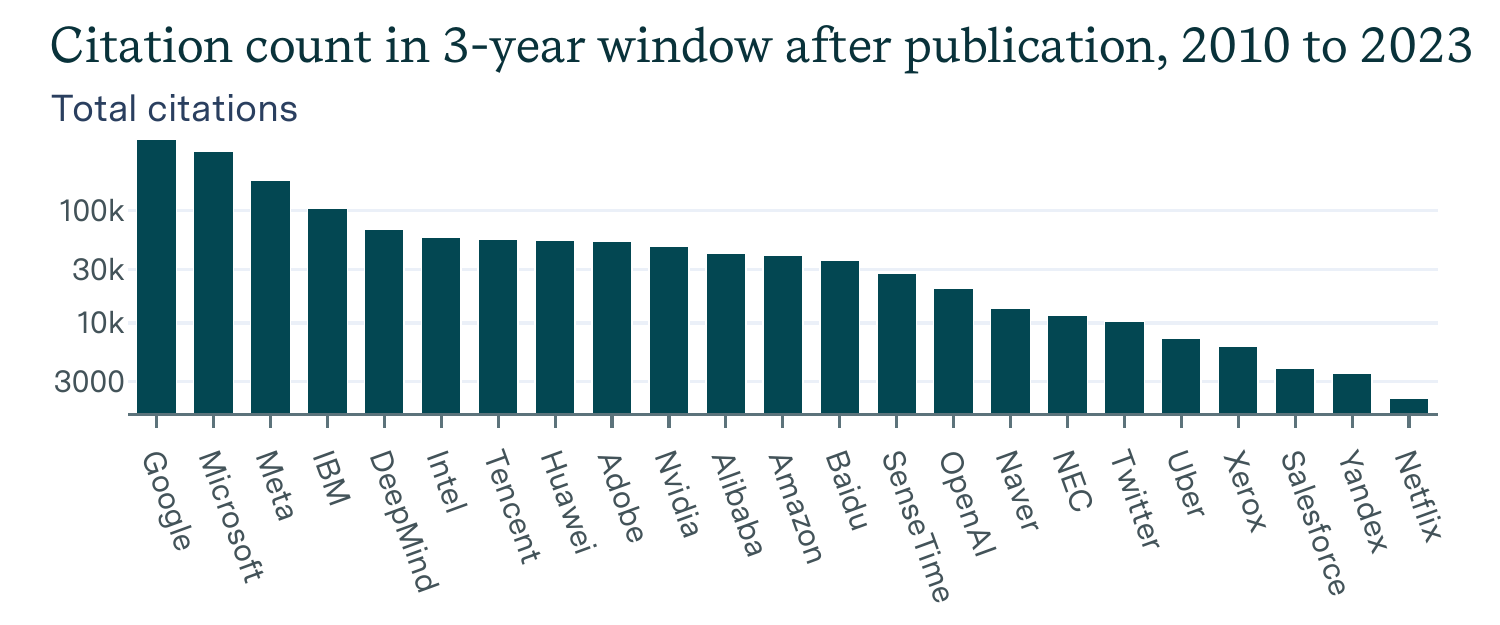}
        \caption{\centering \footnotesize Number of publications and citations for AI/ML publications between 2010 to 2023 by company}
        \label{fig:publications_citations}
    \end{subfigure}
    ~
    \begin{subfigure}{0.49\linewidth}
        \includegraphics[width=\textwidth]{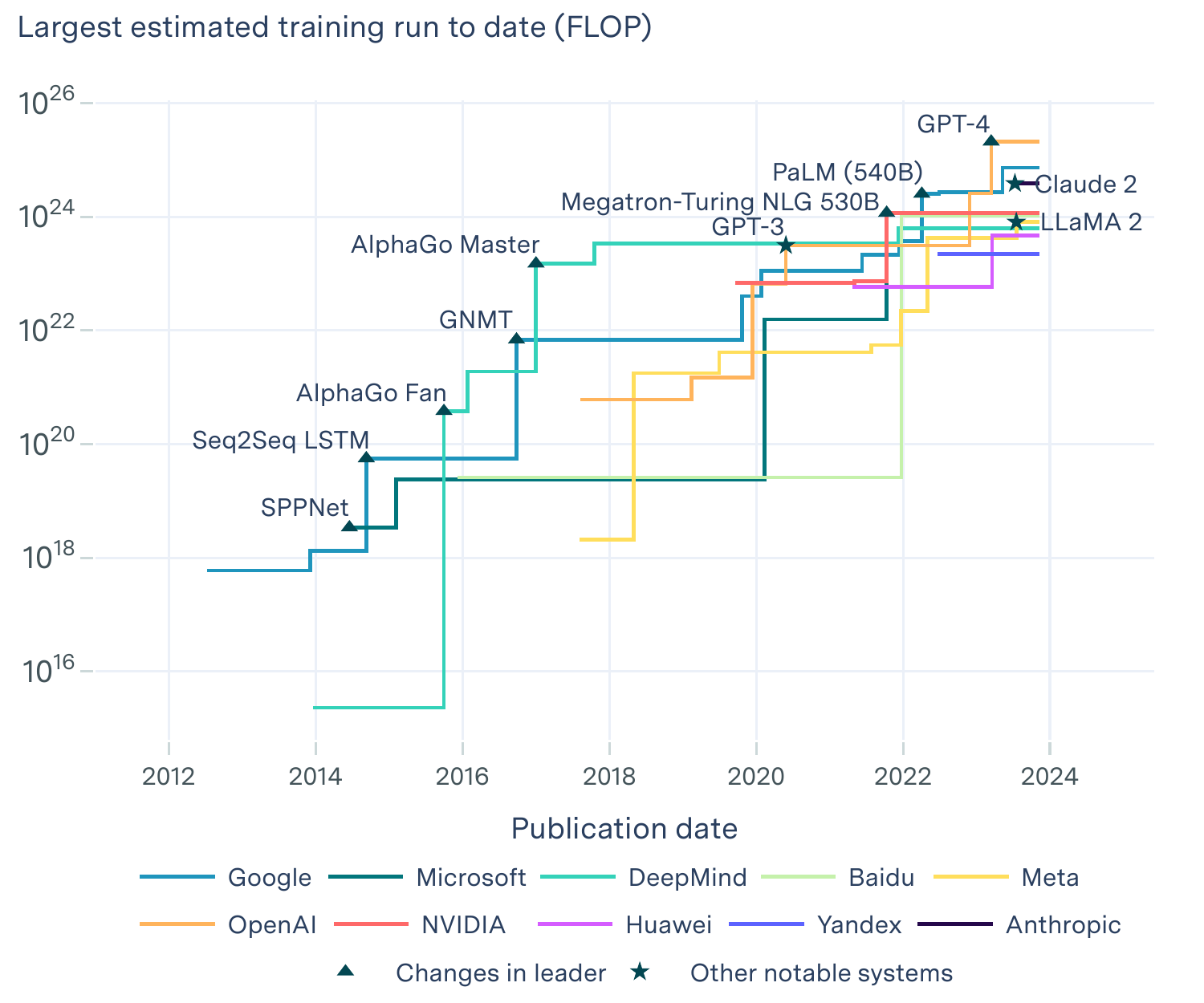}
        \caption{\centering \footnotesize Largest publicly announced \ac{AI} training runs by company over time}
        \label{fig:companies_largest_compute}
    \end{subfigure}
    
    \begin{subfigure}{0.49\linewidth}
        \includegraphics[width=\textwidth]{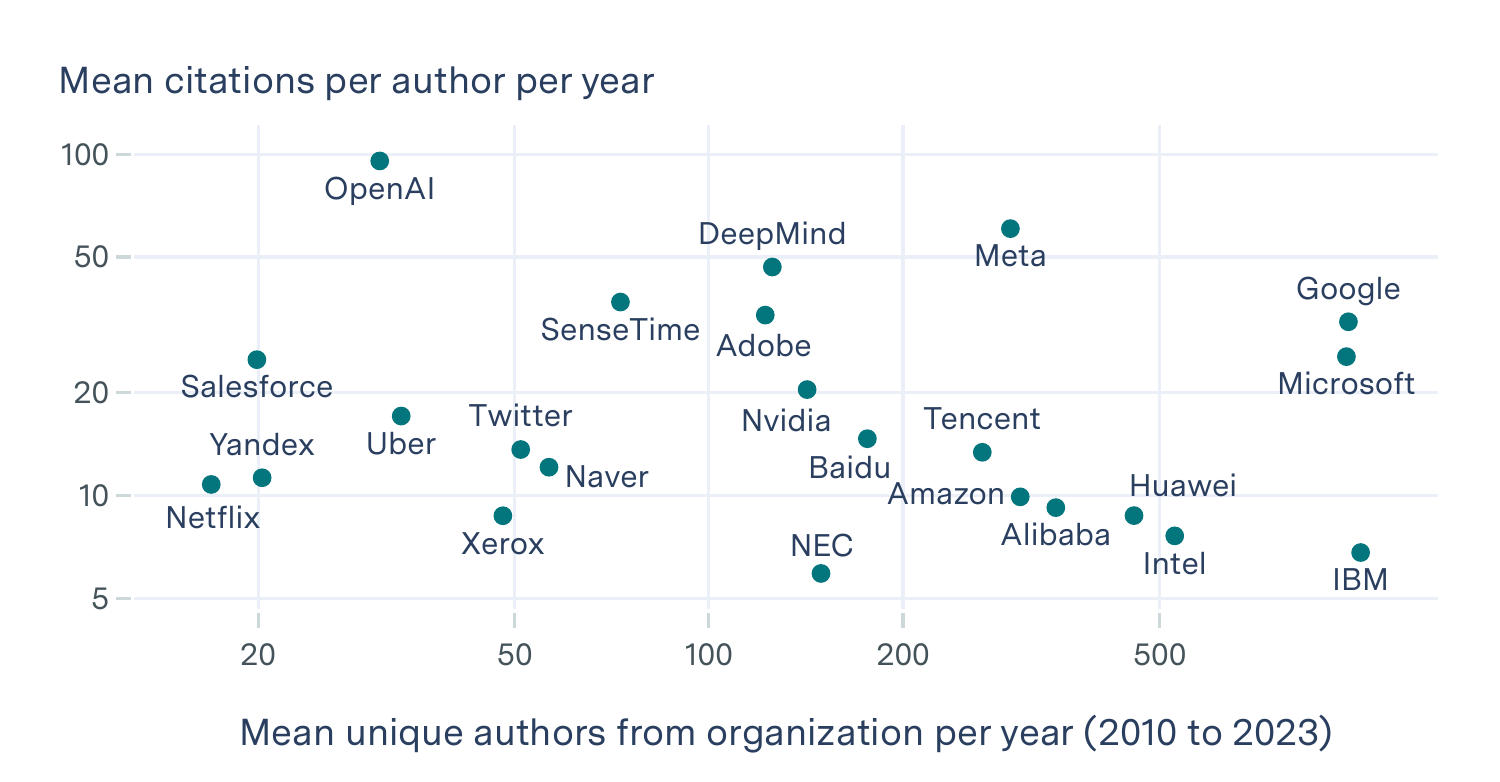}
        \caption{\centering \footnotesize Citations per year per author and mean number of unique authors for AI/ML publications, by company.}
        \label{fig:citations_authors}
    \end{subfigure}
    ~
    \begin{subfigure}{0.49\linewidth}
        \includegraphics[width=\textwidth]{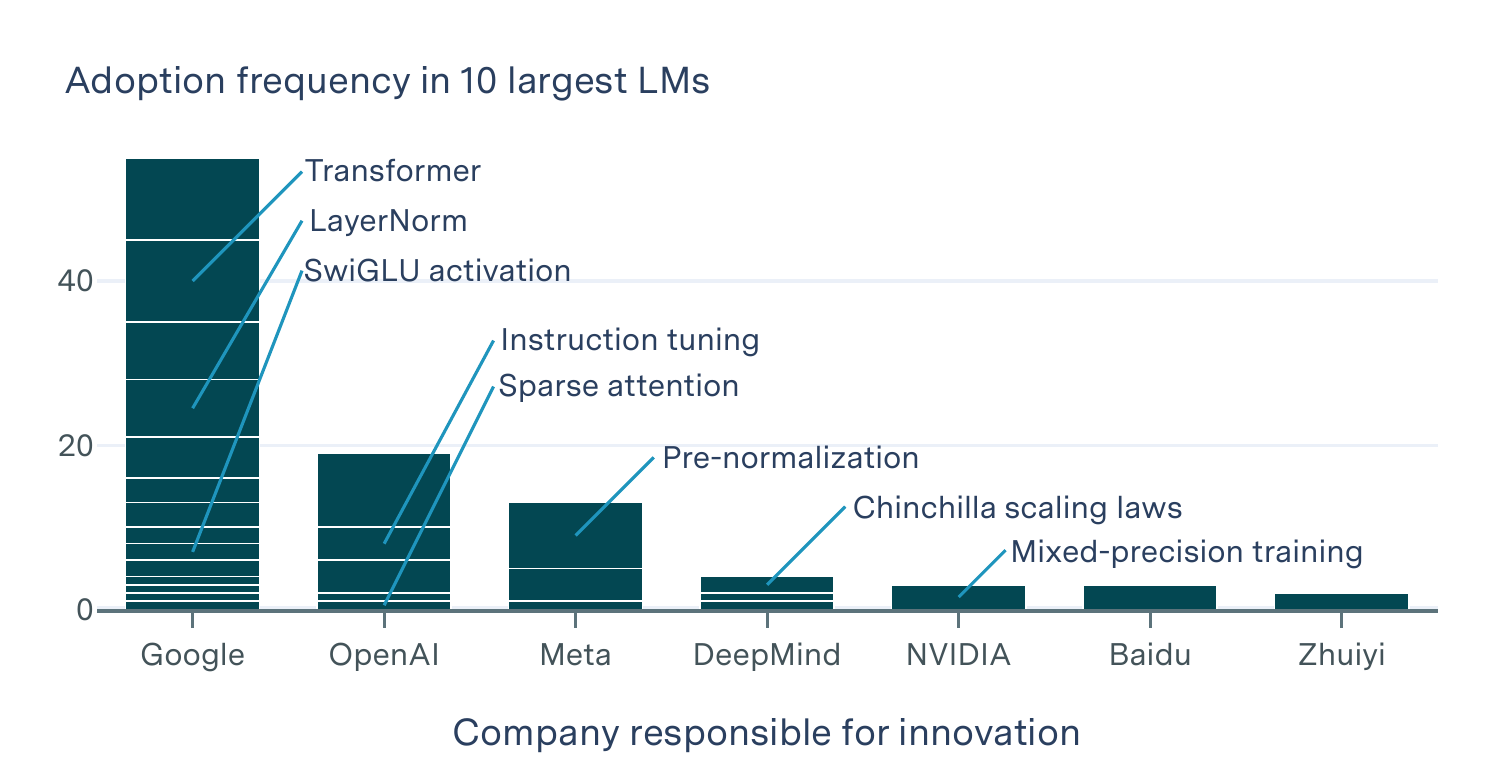}
        \caption{\centering \footnotesize Adoption frequency of innovations underpinning the largest \ac{LMs}, by company responsible for the innovation}
        \label{fig:num_innovations}
    \end{subfigure}
    
    \caption{\footnotesize Key metrics for leading companies in \ac{AI} research. In all metrics, ``Google'' aggregates Google AI labs except for DeepMind. (a) Total number of publications and citations. (b) The largest training run reported in a publication for companies that lead in \ac{AI} scaling, based on the \ac{PCD} database. (c) Mean citations per author versus the mean number of authors publishing papers. Each value is an arithmetic mean over the years 2010 to 2023. (d) Adoption frequency of innovations underpinning the ten largest publicly documented \ac{LMs} in terms of training compute, grouped by company responsible for the innovation. Full methodology can be found in \textit{Appendix}.}
    \label{fig:key_plots}
\end{figure*}

We find that Google and Microsoft have led in terms of total publications and citations over the past 13 years (Fig. \ref{fig:publications_citations}). OpenAI, Meta and DeepMind\footnote{In 2023, DeepMind merged with Google Brain to form Google DeepMind. We separate DeepMind because historically it had significant operational independence from parent company Alphabet.} accumulated the most citations relative to their number of publications, higher than other industry leaders such as Microsoft and other AI labs at Google; see Fig. \ref{fig:citations_authors}. In contrast, publications from Chinese companies, such as Tencent, Baidu and Huawei received fewer citations, especially in per-author terms, indicating that Chinese companies have notably lower citation-impact per researcher. SenseTime is a notable exception, achieving 37 citations-per-author, above Microsoft (26) and below DeepMind (47). These comparisons hold across earlier and later time periods (see \textit{Appendix}).

Advances in \ac{ML} have also been associated with increases in training compute, as measured in \ac{FLOP} \cite{sevilla2022compute}. Training compute thus provides a complementary perspective on which companies are leading in \ac{AI}, and how this lead has changed. We find that several leading US companies have followed a similar strategy of rapidly scaling their training compute (see Fig. \ref{fig:companies_largest_compute}).
Google's most compute-intensive, publicly announced training run grew approximately 10 million-fold over 11 years, while OpenAI and Meta grew by approximately one million-fold in six years. By contrast, for notable \ac{ML} systems in general, the average compute budget grew approximately four thousand-fold in six years \cite{sevilla2022compute}.

The company responsible for the most compute-intensive training runs has undergone significant change from 2012 to 2023 (see Fig. \ref{fig:companies_largest_compute}). Our data indicates that Google's \ac{AI} labs led compute scaling for much of the 2010s. However, OpenAI, founded in 2015, proceeded to rapidly match the frontier of training compute established by DeepMind's AlphaGo systems. To the best of our knowledge, OpenAI now holds the record for the largest training run with GPT-4 \cite{openai2023gpt4}, which used an estimated $2\mathrm{e}25$ FLOP \cite{epochMachineLearningData2022}. OpenAI's rapid scaling has been assisted by a partnership with Microsoft announced in 2019 \cite{openaimicrosoft2019}. Anthropic is also notable for Claude 2 at an estimated $4\mathrm{e}24$ FLOP, which is close to Google's PaLM 2 ($7\mathrm{e}24$ FLOP). This is despite Anthropic emerging too recently for our publication metrics. The largest publicly disclosed training runs from Google, Meta, and Baidu are estimated to be within a factor of 100-fold of the frontier set by GPT-4. However, Meta has consistently trailed behind the frontier, with the smallest gap estimated at 6-fold in May 2022.

In light of the crucial role of large language models such as PaLM and LLaMA in contemporary \ac{AI} research, we also assessed the adoption of algorithmic innovations underpinning the largest 10 such models that are publicly documented, as measured by quantity of training compute. Since algorithmic innovations represent a fundamental contribution of \ac{ML} research, quantifying how often they are adopted provides a further useful perspective on research impact.

We find that innovations affiliated with Google (excluding DeepMind)---such as the Transformer \cite{vaswani2017attention} and LayerNorm \cite{ba2016layer}---are the most frequently adopted in recent foundation models (Fig. \ref{fig:num_innovations}). OpenAI's innovations have been adopted roughly one-third as much as Google's, with key innovations in prompting for in-context learning \cite{brown2020language}, and instruction tuning \cite{ouyang2022training}. OpenAI is followed by Meta and DeepMind.

\section*{Discussion}\label{sec:discussion}

Overall, our results quantify the particular importance of Google, OpenAI, and Meta in industry \ac{AI} research, as these companies were especially prolific across all the metrics---innovations, compute, and citation-impact. Especially notable is their dominance in recent innovations that underpin large \ac{LMs}, as they were responsible for most of the innovations we identified.

Furthermore, our metrics can be used to compare companies of different size and age. Dividing citations by unique authors quantifies the large impact of research labs like OpenAI and DeepMind, relative to their size. In addition, looking at the largest training runs and the adoption of recent innovation in large \ac{LMs} reveals which companies have pushed the frontier of capabilities. These metrics are therefore an important complement to more general bibliometrics when assessing key players in the \ac{AI} industry. Expanding the innovation data to other domains of \ac{AI} research is a valuable direction for future work. 

Our metrics suggest that the top Chinese \ac{AI} labs substantially trail their US peers in research impact. Except for SenseTime, Chinese companies received far fewer citations per author than Google and Microsoft. This is despite high publication counts from companies like Huawei and Tencent, ranking 4th and 6th in total publication count respectively. These differences suggest a potential talent gap between Chinese and US industry \ac{AI} labs in the traditionally academic venues covered by OpenAlex data. However, note that a large portion of Chinese \ac{AI} research activity is not in such venues, and not captured by our metrics. For instance, Tencent has been assigned more patents since 2010 than Microsoft and Google \cite{cset-cat}. Recent trends also suggest that Chinese industry could quickly scale up compute investments (see the discussion of Fig. \ref{fig:companies_largest_compute} in Results). However, it remains uncertain whether scaling would be enough to compensate for other potential discrepancies, such as talent. 

As well as quantifying research impact, our results highlight the rapid scaling of compute among \ac{AI} industry leaders. Fig. \ref{fig:companies_largest_compute} illustrates how rapidly OpenAI caught up to Google's \ac{AI} labs. Anthropic is yet another example that only emerged in 2021---albeit with a large number of former OpenAI employees. The training compute of Anthropic's Claude 2 system is estimated to be $4\mathrm{e}24$ FLOP \cite{epochMachineLearningData2022}, which would make it the third largest ever training run to date.

As governments increase oversight of AI companies, this work provides an empirical basis to inform policy discussions. Our quantification of publication impact, training compute, and innovations establishes evidence of research contributions and rapid scaling by leading firms. Policymakers shaping AI governance must account for the capabilities and growth of frontier companies.

\section*{Materials and methods}

\subsection*{Publication dataset}

We sourced all publication data from OpenAlex \cite{priem2022openalex}.
First, we narrowed down to a set of top 25 companies based on affiliations with (1) the top-100,000 most cited \ac{AI} and \ac{ML} works, and (2) notable \ac{ML} systems according to the \ac{PCD} database.
We used keyword matching to merge entities associated with the same company. We then obtained all \ac{AI} and \ac{ML} works (including pre-prints) in OpenAlex that had at least one affiliation in the top 25 companies, and which were published between 2010-01-01 and 2023-06-01. Two companies ranked in the top 25 due to a single outlier and were therefore excluded. Due to the unusually small amount of annotated data retrieved for OpenAI, we identified and manually annotated 82 additional works (see \textit{Appendix} for details). The final dataset contained 66,175 unique publications.

\subsection*{Citations and authors}

To control for publication age, we counted citations in the publication year and the three subsequent years, rather than the total citations to date. Our conclusions are not sensitive to this choice (see \textit{Appendix}).
The unique author counts were obtained from author affiliation data in the publication dataset.
For the $y$-axis of Fig. \ref{fig:citations_authors}, we divided the truncated citation count by the number of unique authors in each year, and then averaged over the years.
In cases where multiple companies were affiliated with the same publication, we attributed the citation count to all of those companies, and attributed each author to their respective affiliation.

\subsection*{Training compute}

We only considered the training compute of \ac{ML} systems that are recorded in the \ac{PCD} database.
The ten companies in Fig. \ref{fig:companies_largest_compute} were selected for having the top ten largest publicly announced training runs as of June 1st, 2023. The training compute values are estimates \cite{sevilla2022compute}.

\subsection*{Algorithmic innovations}

We identified innovations based on a survey of large language models \cite{zhao2023survey}. We included innovations that apply to language model pre-training, including variations of the Transformer architecture. We also included post-training or fine-tuning techniques, such as instruction tuning \cite{ouyang2022training}.
We then identified the origin and affiliations of each innovation.
The origin was the earliest publication (including pre-prints) that introduced the innovation and was in the citation tree of the survey paper or one of the chosen large language model papers.
Finally, for each of the top 10 largest \ac{LMs} that were publicly documented before the cutoff date of June 1st, 2023, we recorded whether each innovation was adopted based on evidence in the paper.

\section*{Acknowledgements}

We thank Lennart Heim, Ross Gruetzemacher, David Manheim, Anson Ho, Helen Toner, Zachary Arnold, James Dunham, and Michael Aird for helpful comments.

\bibliography{references.bib}

\appendix

\section*{Appendix: Extended Methods}

Code and instructions to download the datasets are available at \url{https://github.com/epoch-research/ai-research-impact}.

\subsection*{Merging entities}

As noted in \textit{Materials and Methods}, we merged entities in OpenAlex that were associated with the same company. Institutional affiliations in OpenAlex are somewhat specific; for example, ``Microsoft (United States)'' is distinct from ``Microsoft Research (United Kingdom)''. Distinguishing these entities can be useful, as some will have greater impact than others. However, we merged these affiliations for two reasons. Firstly, the fine-grained institutional affiliations can be inaccurate. For instance, the publication ``Densely Connected Convolutional Networks'' (OpenAlex ID: W2963446712) has an affiliation labeled as ``Meta (Israel)''. However, the actual authorship in the paper is Laurens van der Maaten, who was affiliated with Facebook AI Research, and to our knowledge was not affiliated with Meta's Israel offices in particular. The second reason is that certain decisions are made at the company level---such as the allocation of R\&D budget and cloud compute---which greatly impact the activities of individual labs.

\subsection*{Selecting companies}

To make it tractable to rank companies using OpenAlex data, we first narrowed down to a set of top 25 companies based on affiliations with (1) the top-100,000 most cited works assigned with the “Artificial Intelligence” concept or “Machine Learning” concept in OpenAlex, and (2) notable ML systems according to the \ac{PCD}
database.
We used both of these data sources because they provide different perspectives on industry \ac{AI} leadership. Going by total citations alone, a company such as OpenAI would be excluded due to its small size. Going by notable \ac{ML} systems alone, a company such as SenseTime would be excluded due to not publicising many of its \ac{ML} systems.

For the top-cited works, we credited citation counts to companies using the same method as the final results (see \textit{Measuring citations}). For notable ML systems, we counted the number of systems associated with each institution. After obtaining citation and notable system counts, we normalized each set of counts to standard scores, and added those two standard scores together for each institution. Institutions that appeared in the citation data but not in the notable systems data were assigned a notable systems count of zero before being normalized. The 25 companies in the publications dataset are the top 25 companies according to the sum of the two standard scores. We chose 25 companies as the cutoff somewhat arbitrarily. One reason is that the scores rapidly drop off to a long tail past the top 5, so companies beyond the top 5 are less differentiated. Another reason is to make our analysis easier to follow, as fewer companies need to be compared.

\subsection*{Creating the publications dataset}

We constructed the dataset of publications affiliated with the 25 selected companies by querying works in OpenAlex for which (1) the authorship had at least one affiliation in the list of selected companies; (2) the work was tagged with the ``Artificial Intelligence'' concept or the ``Machine Learning'' concept (concepts in OpenAlex are similar to research fields); (3) the work was published between 2010-01-01 and 2023-06-01 inclusive. Once we amended some of the data affiliated with OpenAI (see \textit{Amendments to OpenAI data}), we obtained the final publications dataset.

\subsection*{Reduced data availability since 2021}

In preliminary analysis, we found that OpenAlex had less publication data available in 2022 and (to a lesser extent) 2021, and that this was due to one of OpenAlex's sources, Microsoft Academic Graph, no longer being updated with new entities as of May 2021.\footnote{See \url{https://www.microsoft.com/en-us/research/project/academic/articles/microsoft-academic-to-expand-horizons-with-community-driven-approach/}}
When aggregating metrics across all years, we still used all of the data up to the 2023 cutoff. This systematically reduced the measured number of publications, citations etc. relative to the true values. Note however that our high-level conclusions hold for earlier time periods; see \textit{Measuring the variation of publication-based results over time}.

\subsection*{Deduplication}

When fetching works from OpenAlex, any collaboration between our selected companies can cause duplication of works in the dataset. We only counted one of these duplicates per company in our results. Additionally, we removed works that had different IDs but identical titles, after filtering out non-alphanumeric characters from the title. These duplicates can occur due to similar papers being published in multiple venues, or the same paper being hosted on multiple websites. We found that such duplicates made up 14\% of the works in our initial list. To avoid any doubt about double-counting citations, we only kept the duplicate which had the highest citation count, removing the others. If the citation counts were equal, we chose the work with the lower ID number.

We also de-duplicated authors. For each company in each publication year, we removed any authors whose exact \verb+display_name+ already appeared under some other author ID. Overall, approximately 2\% of all authors were duplicates in this sense, with higher percentages for Chinese companies (3 to 6\%). This assumed that the IDs in question were duplicated in error. In some cases, this process may have removed a distinct author who happened to have an identical name to someone else at the same company in the same year. We assumed this was less likely than duplication, and therefore that the de-duplication process was a net improvement in data quality. There is still likely to be some degree of duplication in authors after this process, due to e.g. an author's use of initials in some papers but not others.

\subsection*{Accuracy of concept labelling}

OpenAlex labels each work with a list of ``concepts'', similar to topics or fields of research. We filtered publications that were labeled with ``Artificial Intelligence'' or ``Machine Learning'' to obtain our publications dataset.  Concepts are organised into a hierarchy, so filtering by the level 1 concept of ``Artificial Intelligence'' will include works with the level 2 concept of ``Deep Learning''. Note that ``Artificial Intelligence'' and ``Machine Learning'' are both level 1 concepts in OpenAlex.

To check the accuracy of concept labelling in OpenAlex, we conducted a cursory analysis of both false negative and false positive rates. For false negatives, we used the \ac{PCD} database \cite{epochMachineLearningData2022}. Each \ac{ML} system in this database includes the title of the source publication, which we used as a ground truth for a ``machine learning'' publication. At the time of writing there were 818 entries in the database. We limited the set of publications to arXiv papers because this was the easiest to check. We also filtered for works which had an affiliation with one of our 25 selected companies. This resulted in 128 unique matching works in OpenAlex. Of these 128 works, 125 were in our final dataset of publications---a 2.3\% false negative rate. This is probably lower than the overall false negative rate, because both OpenAlex and the \ac{PCD} database  are more likely to document highly-cited, easily-accessed works. However, this result shows that one set of important publications is well represented in our dataset.

To get a sense of false positive rates, we compared our publications dataset with the Experimental AI corpus, a dataset derived from OpenAlex \cite{mateos2022experimental}. This dataset uses more rigorous filtering to remove false positives from the raw OpenAlex data. Of the 100 most-cited works in our dataset published in 2021 or earlier, only 61 were in the Experimental AI corpus. However, upon inspection only four of the missing 39 works seem questionable (e.g. ``Accurate, Dense, and Robust Multiview Stereopsis''; ``SciPy 1.0: fundamental algorithms for scientific computing in Python'').

\subsection*{Accuracy of institutional affiliations}

To check how accurate OpenAlex was at labeling institutional affiliations, we examined how often the \verb+raw_affiliation_string+ field of each company's works matched some keywords. If none of the raw strings matched the keywords, we called this a ``non-match''.
We found that the highest non-match rate was 22\% (for the company Uber), while the average was 9\%.
However, these non-match rates are loose upper bounds on the actual false-positive rate of affiliation labels. 
The main reason for this is that OpenAlex does not only use the \verb+raw_affiliation_string+ to determine affiliation.
For most non-matches, the \verb+raw_affiliation_string+ is simply empty.
For three works with an empty \verb+raw_affiliation_string+ that we manually inspected, we found the affiliation label to be correct.\footnote{The IDs for the three works were W2963403868, W2470673105, and W2970971581. These were the first works fetched for Google, Microsoft and Meta for our publication dataset.}
In other cases, our keywords failed to pick up on relevant information in \verb+raw_affiliation_string+. For example, one string said ``T.J. Watson Research Center'' without reference to IBM, but was correctly labeled by OpenAlex as ``IBM (United States)''.

While inspecting \verb+raw_affiliation_string+ data, we discovered that OpenAlex systematically mislabeled affiliations with SenseTime as ``Group Sense''. Approximately 98\% of the works labeled as ``Group Sense'' had some variation of ``SenseTime'' in the \verb+raw_affiliation_string+. Given how consistent this mislabeling was, we simply renamed  ``Group Sense'' to ``SenseTime'' in our results.

The most notable example of incorrect labeling was a version of the AlexNet paper published in Communications of The ACM in 2017 (OpenAlex ID: W2618530766). This was one of the most cited publications in our dataset, with 18,199 citations at the time of writing. OpenAlex affiliates this publication with OpenAI because Ilya Sutskever was at OpenAI by 2017. However, this was not the case for the original AlexNet publication. Furthermore, the labeling is mismatched such that Ilya Sutskever is affiliated with Google while Geoffrey Hinton is affiliated with OpenAI. Since this was a particularly impactful outlier, and the original AlexNet paper did not have industry affiliations \cite{krizhevsky2012imagenet}, we manually removed the 2017 publication from our dataset.

An example of a false negative was the GPT-3 paper \cite{brown2020language}. Initially, the authorship data in OpenAlex made no reference to OpenAI, even in the \verb+raw_affiliation_string+. However, this was partially corrected by the amendments we made to OpenAI data, detailed in the next section.

\subsection*{Amendments to OpenAI data}

OpenAI was particularly disadvantaged among the selected companies in terms of labeling in OpenAlex. Filtering works by \verb+raw_affiliation_string+s that include the terms ``openai'' or ``open ai'' yielded 236 results, while filtering works by institutional affiliation with OpenAI yielded only 81 results. The relative difference in those counts was unusually large. By contrast, the same comparison for DeepMind (using ``deepmind'' as the search term) yielded 1182 results vs. 801 results. Since OpenAI had the smallest publication count of all the selected companies, the citation and author metrics for OpenAI were also particularly sensitive to outliers and errors (an example of both an outlier and error was the 2017 AlexNet paper discussed previously).

For these reasons, we manually amended OpenAI publication data in our dataset. To do this, we first scraped titles of research articles from the OpenAI research page (\url{https://openai.com/research}). We then matched these titles to works in OpenAlex, using a threshold value for string similarity. Of those works, we manually added an institutional affiliation with OpenAI for authors which had ``OpenAI'' in the \verb+raw_affiliation_string+, or which were affiliated with OpenAI on a different paper in our dataset. Finally, we replaced any works in the original dataset that overlapped with this new set, so that they contained the new affiliation labels. 

This process resulted in 82 additional OpenAI-affiliated works that were not previously in the publication dataset, bringing the total OpenAI-affiliated works to 163---still the smallest of the 23 companies included in the results. The additional works increased OpenAI's total citation count from approximately 13k to 20k, which only moved OpenAI up one rank in total citations. The change also \emph{decreased} OpenAI's citations per author per year from 163 to 96, because its annual average unique author count increased from to 16 to 31.

\subsection*{Measuring citations} \label{citations}

To control for publication age, we counted citations for each work in its publication year and the three subsequent years, rather than the total citations to date. Note that the three-year horizon meant that works published in 2020 onwards had their horizon truncated to the dataset cutoff of June 1st, 2023, and therefore received fewer citations than otherwise. We decided that three years was the best compromise between (a) allowing more time for a publication to have impact, and (b) having more years of data before the citation horizon cuts off. Citation counts before 2012 were unavailable in OpenAlex, so citation counts for works published in 2010 and 2011 were also underestimates. In the main text we report the sum of the truncated citation count for all publications affiliated with each company (which were published between 2010 and June 2023).

We assigned the same credit for publication and citations to all of the companies affiliated with a given publication. For example, suppose a publication has 32 citations, and one author is affiliated with Microsoft while another is affiliated with Nvidia. In this case we would have assigned a value of 1 each to the publication counts of Microsoft and NVIDIA, and a value of 32 each to the citation counts of Microsoft and NVIDIA. However, it is worth noting that only 3.6\% of publications in our dataset were affiliated with more than one of the 25 selected companies. Of those, only 5\%  involved more than two selected companies. So the choice of credit assignment only has a small effect on our results.

In preliminary experiments, we tested the effect of the citation count horizon on our final results. We found that our high-level conclusions were robust to this change. Using a 1-year horizon (i.e. counting citations in the publication year and the following year), the top five companies by total citation count were identical. The largest shifts in ranking were Intel moving from 6th to 8th and Alibaba moving from 11th to 9th. For citations per author, there was slightly less spread between companies, with values ranging on the order of 2 to 20 rather than 5 to 100. Meta decreased in citations per author relative to OpenAI, becoming similar to DeepMind.

With no truncation of citation count (equivalent to a 13-year horizon for our dataset), the ranking of the top six companies by total citations was identical. However, Microsoft was no longer significantly different from Google. Tencent moved from 7th to 9th, swapping places with Adobe, while Baidu moved from 13th to 11th. For citations per author, there was greater overall spread between companies, with values ranging on the order of 20 to 200. Meta became comparable to OpenAI on this metric, while DeepMind pulled further ahead of SenseTime. Baidu also pulled further ahead of the remaining Chinese companies, all of which still scored far lower than the leading US companies.

\subsection*{Disagreement with Semantic Scholar citation counts}

We found that OpenAlex citation counts were nearly 2x lower (on average) than corresponding citation counts in Semantic Scholar, which is a popular source of bibliometric data.\footnote{\url{https://www.semanticscholar.org/}} In this section we describe how we measured the agreement between these two sources, and why we decided to use citation counts reported by OpenAlex.

To assess the agreement of OpenAlex citation counts with Semantic Scholar, we compared a subset of our publication dataset to corresponding entries in Semantic Scholar at the time of writing, October 31, 2023. We used DOIs to match the OpenAlex publications with Semantic Scholar entries. Not all 66,175 papers could be matched, leaving 42,255 publications.

For this sample we found that Semantic Scholar reports approximately 1.9x higher citation counts than OpenAlex on average (median 1.4x, standard deviation 2.5), excluding OpenAlex publications with zero citations. 
If Semantic Scholar citation counts are more accurate, this means that OpenAlex systematically underestimates citations. Some of the OpenAlex data may have been outdated due to the retirement of one of its main sources, Microsoft Academic Graph, in May 2021. However, this is a small amount of time relative to the 2010--2023 range of our dataset, so it seems insufficient to explain the entire difference of 1.9x. The citation-counting methods used by Semantic Scholar may simply be more effective than OpenAlex. However, it is also possible that Semantic Scholar overestimates citations e.g. due to AI-powered methods that are sensitive to false positives, or due to counting the same citation from multiple sources.

The level of agreement in citation counts also varied by company. The least agreement occurred for OpenAI (4.2x lower than Semantic Scholar), DeepMind (3.6x), Salesforce (2.7x), Uber (2.4x), and Amazon (2.3x). The most agreement occurred for Xerox (1.4x), Intel (1.5x), NEC (1.5x), Twitter (1.5x), IBM (1.6x), and Netflix (1.7x). Other companies had similar levels of agreement to the average: Meta (1.9x), Google (2.1x), Microsoft (2.0x), Adobe (2.0x), Huawei (1.9x), Baidu (2.0x), Nvidia (2.1x), Yandex (1.9x), Tencent (2.1x), Naver (2.1x), Alibaba (2.1x), and SenseTime (2.1x).
Since our primary goal was to compare companies leading in \ac{AI} research on a relative basis, the overall underestimation of citation counts in OpenAlex was acceptable for our purposes.

The difference in agreement for each company was more problematic, because it suggests a potential bias towards some companies. However, our overall conclusions would not change if we adjusted for these biases using the above ratios: OpenAI and DeepMind would be even more favoured, while the Chinese companies in the set would be no less favoured than Meta, Google and Microsoft. Given this, and given the uncertainty about which database is more accurate, we decided to keep using OpenAlex citation counts in this work.

\subsection*{Measuring unique authors}

We measured the number of unique authors on works published in each year, affiliated with each company. We then calculated the mean number of unique authors over the years 2010 to 2023. To measure citations per author, we divided the total citation count (as described in \textit{Measuring citations}) for each year by the total unique authors in that year, and then averaged over the years 2010 to 2023.

\subsection*{Statistical significance of citation count differences}

We performed two-sided Mann-Whitney U-tests to compare the distribution of citations for each company in the whole time period of 2010--2023. Mean citations for OpenAI and DeepMind were 125 and 91, and the distributions of citations differed significantly ($U=82856.5, n_1=163, n_2=755, p<0.001$). The citation distributions for DeepMind and SenseTime did not differ significantly ($p=0.08$). Meta and Google had means of 56 and 43, differing significantly ($U=17509800, n_1=3329, n_2=10094, p<0.001$). Microsoft had a mean of 23 while Baidu had a mean of 19, differing significantly ($U=77623307, n_1=14550, n_2=1920, p<0.001$). Baidu had a similar distribution to Tencent ($p=0.94$) but differed significantly to Alibaba ($U=3425506.5, n_1=1920, n_2=3424, p=0.01$).

\subsection*{Measuring the variation of publication-based results over time} \label{persistence}

To check the robustness of the publication-based results, we broke down the publication and citation counts of companies into the periods of 2010--2016 and 2017--2023. In both periods, Google and Microsoft still had the highest citation counts. However, IBM had a higher publication count than Google in the 2010--2016 period, placing second behind Microsoft. In the same period, US companies Google, Microsoft and Meta were still ahead of Chinese companies Huawei, Baidu, Alibaba and Tencent.

For citations per author, there were more noticeable changes but most of our conclusions held. OpenAI and Meta still led on this metric. Google and Microsoft remained ahead of Baidu, Alibaba and Huawai in citations per author in 2017--2023. Notably however, Tencent rose on this metric from less than 10 in 2010--2016 to more than 30 in 2017--2023 to rival Microsoft.

For unique author counts, OpenAlex reported much fewer publishing authors in 2010--2016 compared to 2017--2023. For example, Baidu had an average of over 200 unique authors per year in the 2017--2023 period, but less than 50 in the 2010--2016 period. Meanwhile, Microsoft and IBM maintained a similar number of unique authors across the two time periods.

\subsection*{Training compute}

Our training compute analysis relied on publicly documented AI systems from the \ac{PCD} database \cite{epochMachineLearningData2022}. As such, it may exclude proprietary training runs that remain undisclosed. However, the \ac{PCD} records comprise the most comprehensive public account of training compute over time. The estimates also undergo continual refinement as new information emerges \cite{epoch2023announcingupdatedpcddatabase}. Given the broad coverage and regular updates, we expect any gaps or inaccuracies in training milestones to be relatively minor and unlikely to affect the overall leadership trends observed. But due to reliance on public data, our findings regarding compute scaling leadership remain subject to uncertainty from potential undisclosed private advances.

\subsection*{Algorithmic innovations}

\subsubsection*{Selection of innovations and models}

We limited our focus to key algorithmic innovations in large \ac{LMs} because (a) this is arguably the paradigm driving the most impressive capabilities today, (b) assessing all innovations in \ac{ML} was too large in scope. Our method provides an example that we hope future work will further develop and expand.

We identified the set of innovations based on the survey of large language models by \cite{zhao2023survey}.
Since the focus was on industry contributions, our results excluded innovations which only had academic affiliations at the time of publication, e.g. the Adam optimizer \cite{kingma2017adam}.
Data on those innovations can still be found in the full dataset.
The full list of 26 innovations, which had industry affiliation and were adopted at least once in our set of \ac{LMs}, is as follows: Kaplan et al. scaling laws \cite{kaplan2020scaling}, Hoffmann et al. (i.e. Chinchilla) scaling laws \cite{hoffmann2022training}, Transformer (general architecture) \cite{vaswani2017attention}, sparse attention \cite{child2019generating}, multi-query attention \cite{shazeer2019fast}, LayerNorm \cite{ba2016layer}, pre-normalization \cite{baevski2019adaptive}, learnable position embeddings \cite{gehring2017convolutional}, sinusoidal position embeddings \cite{vaswani2017attention}, relative position embeddings \cite{dai2019transformerxl}, rotary position embeddings \cite{su2022roformer}, SwiGLU activation \cite{shazeer2020glu}, Sparsely-Gated Mixture-of-Experts layer (MoE) \cite{shazeer2017outrageously}, encoder-decoder Transformer \cite{vaswani2017attention}, causal decoder Transformer \cite{liu2018generating}, language modeling task (with Transformer architecture) \cite{liu2018generating}, cloze task (with Transformer architecture) \cite{devlin2019bert}, denoising autoencoding task (with Transformer architecture) \cite{lewis2019bart}, dynamic batch size \cite{smith2018dont}, Adafactor optimizer \cite{shazeer2018adafactor}, mixed precision training \cite{micikevicius2018mixed}, instruction tuning \cite{ouyang2022training}, reinforcement learning from human feedback (RLHF) \cite{christiano2023deep}, asynchronous actor-critic (A2C) \cite{mnih2016asynchronous}, prompting for in-context learning \cite{brown2020language}, and chain-of-thought \cite{wei2022chain}. See the dataset for further information.

The largest language models were measured by training compute estimates from the PCD database \cite{epochMachineLearningData2022}. The 10 models for which we measured adoption frequency are as follows: PaLM 540B \cite{chowdhery2022palm}, Megatron-Turing NLG 530B \cite{smith2022using}, ERNIE 3.0 Titan \cite{wang2021ernie}, Gopher 280B \cite{rae2022scaling}, Chinchilla 70B \cite{hoffmann2022training}, PanGu-$\Sigma$ \cite{ren2023pangusigma}, LLaMA 65B \cite{touvron2023llama}, OPT-175B \cite{zhang2022opt}, Yuan 1.0 \cite{wu2021yuan}, and AlphaCode \cite{Li_2022}. To our knowledge, and as of our publication cutoff of June 2023, these are the 10 largest pre-trained language models by training compute \emph{excluding} GPT-4, PaLM-2, and GPT-3.5. We excluded these three models due to a lack of public information about their use of the innovations. 

We expect that some innovations we identified are more controversial, e.g. ``language modeling task (with Transformer architecture)'' or ``prompting for in-context learning''. While language modeling is nothing novel, the combination of language modeling with the Transformer architecture was a non-trivial step \cite{liu2018generating}. As for ``prompting for in-context learning'', we deemed it a key innovation because designing prompts that elicit the in-context learning behaviour of large language models was a non-trivial idea, as demonstrated by \cite{brown2020language}. We made a similar assessment of ``chain-of-thought [prompting]'' as an innovation. If we removed ``prompting for in-context learning'' from the dataset, then Meta and OpenAI would swap in rank. Nonetheless, Google and OpenAI would remain leading contributors across all of our metrics.

\subsubsection*{Identifying the origin of innovations}

As mentioned in the main text, we identified the origin of each innovation as the earliest publication (including pre-prints) that (a) ostensibly introduced the innovation, and (b) was in the citation tree of either the survey paper \cite{zhao2023survey} or one of the chosen large language model papers. In some cases this was difficult to determine, as multiple papers introduce similar ideas. For example, ``Pre-normalization'' refers to either (a) normalization before the skip connection in the Transformer architecture, as used by BERT \cite{devlin2019bert} or (b) normalization after the skip connection, as used by Megatron-LM. The Megatron-LM paper found that type (b) was crucial for training stability \cite{shoeybi2020megatronlm}. Both types (a) and (b) are different to the original Transformer architecture where the normalization happens before each layer output, rather than after each layer input \cite{vaswani2017attention}.

The earliest use of pre-normalization for Transformers that we found was \cite{baevski2019adaptive}, which the survey paper also credits. However, the Megatron-LM paper cites GPT-2 \cite{radford2019language} and BERT \cite{devlin2019bert} instead. We did not find explicit mention of pre-normalization in the BERT paper. The GPT-2 paper cites an earlier computer vision paper as inspiration \cite{radford2019language,he2016identity}. However, that paper uses a different model architecture, so we considered it to be a distinct innovation. The most thorough analysis of pre-normalization we found is \cite{xiong2020layer}, but this was only published in 2020. In the end, we chose \cite{baevski2019adaptive} as the origin of pre-normalization due to it being the earliest known application of pre-normalization to the Transformer architecture.

Some innovations were rejected from our sample because their origin could not be determined. For example, weight decay (also known as L2 regularization) is an important regularization method used for training large language models, but its origin could be traced as far back as ridge regression \cite{tikhonov1943stability}. Another example is restarting training from a checkpoint to avoid spikes in training loss. This is mentioned in the survey paper \cite{zhao2023survey} and was used when training PaLM \cite{chowdhery2022palm}. However, we suspect that many other models used this technique during training without reporting it. Given this uncertainty, we excluded the technique from our list of innovations.

\subsubsection*{Counting adoption of innovations}

To measure the impact of innovation, we counted the number of times an innovation was adopted in the largest language models. Adoption frequency better indicates real-world influence than a raw count of innovations, as the value of any given innovation depends on its incorporation into downstream systems.
In practice, we found the distribution of adoption counts largely mirrored the distribution of raw innovation counts. For example, the innovations we credited to Google, OpenAI, and Meta were adopted 55, 19, and 13 times in large language models respectively. The relative values are similar to their 14, 5, and 3 originating innovations.

When counting innovations related to fine-tuning (e.g. ``Instruction tuning'') and inference (e.g. ``Chain-of-thought''), we included variations on the pre-trained models that were developed by the same company.
For example, we counted PaLM 540B and OPT-175B as adopting instruction tuning, due to the existence of FLAN-PaLM \cite{chung2022scaling} and OPT-IML \cite{iyer2023optiml} respectively.

The details above and in previous sections mean that our results on innovations should be interpreted with particular caution. The choices of innovations, their origin, and the \ac{AI} models using them are somewhat arbitrary. There is also uncertainty in adoption frequency, because some research articles are ambiguous about whether a given innovation was used. Our dataset of innovations is just one source of limited evidence, which we intend to improve and expand in future work.

\end{document}